\documentclass[fleqn,twoside]{article}
\usepackage{espcrc2}

\usepackage{graphicx}
\usepackage[figuresright]{rotating}

\newcommand{\AmS}{{\protect\the\textfont2
  A\kern-.1667em\lower.5ex\hbox{M}\kern-.125emS}}

\title{Renormalization-Mass Scale Dependence in QCD Contributions to Semileptonic $b \rightarrow u$ Decay}

\author{M. R. Ahmady\address{Department of Physics, Mount Allison University, Sackville, NB  E4L 1E6, Canada},
        F. A. Chishtie\address{F. R. Newman Laboratory for Elementary-Particle Physics,
	Cornell University, Ithaca, NY  14853, USA},
        V. Elias\address{Perimeter Institute for Theoretical Physics,
	35 King Street North, Waterloo, ON  N2J 2W9, Canada},
	\thanks{Permanent Address: Department of Applied Mathematics,
	University of Western Ontario, London, ON  N6A 5B7, Canada}
	A. H. Fariborz\address{Department of Mathematics/Science,
	State University of New York Institute of Technology, Utica, NY  13504-3050, USA},\\
	D.G.C. McKeon\address[DMP]{Department of Mathematical Physics, National University of Ireland, Galway, Ireland},
	\thanks{Permanent Address:  Department of Applied Mathematics,
	University of Western Ontario, London, ON  N6A 5B7, Canada}
	T. N. Sherry\addressmark[DMP]
        and
        T. G. Steele\address{Department of Physics and Engineering Physics, University of Saskatchewan, Saskatoon, SK  S7N 5E2, Canada}}
       
\begin{document}

\begin{abstract}
QCD contributions to the $b \rightarrow u \ell^- \overline{\nu}_\ell$ decay rate, which are known to two-loop 
order in the $\overline{MS}$ scheme, exhibit sufficient dependence on the renormalization 
mass $\mu$ to compromise phenomenological predictions for inclusive semileptonic 
$B \rightarrow X_u$ processes. Such scale dependence is ameliorated by the 
renormalization-group (RG) extraction and summation of all leading and 
RG-accessible subleading logarithms occurring subsequent to two-loop order 
in the perturbative series. This optimal RG-improvement of the known portion 
of the perturbative series virtually eliminates $\mu$-dependence as a source of 
theoretical uncertainty in the predicted semileptonic $B \rightarrow X_u$ inclusive rate.
\vspace{1pc}
\end{abstract}

% typeset front matter (including abstract)
\maketitle

%\section{FORMAT}

The perturbative QCD decay rate $b \rightarrow  u \ell^- \overline{\nu}_\ell$ has been 
calculated to two-loop (2L) order in the $\overline{MS}$ scheme \cite{TVR99}:
\begin{eqnarray}%1
\Gamma_{2L}/\kappa & = & \left[ m_b (\mu)\right]^5 \left[ 1 + (4.25360 + 5L(\mu))x(\mu) \right.\nonumber\\
& + & (26.7848 + 36.9902L(\mu) \nonumber\\
& + & \left. 17.2917L^2(\mu))x^2(\mu)\right]\nonumber\\
\end{eqnarray}
\begin{eqnarray}%2
x(\mu) \equiv \alpha_s (\mu) / \pi, \; \; L(\mu) \equiv \log(\mu^2 / m_b^2 (\mu)),\nonumber\\
\kappa \equiv G_F^2 |V_{ub}|^2 / 192\pi^3.
\end{eqnarray}
This rate, labelled curve ``2L'' in Figure 1, exhibits substantial dependence on the  
$\overline{MS}$ renormalization scale $\mu$. Such dependence, an $O$(15\%) decline in rate over the range 
$2$ GeV $\leq \mu \leq 9$ GeV, is necessarily a source of theoretical uncertainty.  Moreover, 
the 2L rate does not exhibit any
extremum in $\mu$ identifiable with a point of minimal sensitivity (PMS) \cite{PMS81}, 
although the one-loop (1L) rate, as given by Eq. (1) without its 2L terms, does 
have a maximum near $\mu = 2.7$ GeV at $\Gamma_{1L}/\kappa \cong 1800$ GeV$^5$, as 
evident from the 1L curve in Fig. 1. Such a maximum, however, is no longer 
evident in the 2L curve, whose only benchmark value is its closest approach 
to the 1L curve occurring at $\mu = 2.85$ GeV (and $\Gamma_{2L}/\kappa \cong 1890$ GeV$^5$). We denote 
this point, at which the 2L contribution to the rate is a minimum, as the 
point of fastest apparent convergence (FAC) \cite{GG80}. We further note from 
Figure 1 that the 1L and 2L curves exhibit PMS and FAC points at values 
of $\mu$ much less than the b-quark mass $m_b(m_b) = 4.17$ GeV \cite{KGC81} at which all 
logarithms $L(\mu)$ in Eq. (1) are zero. 

\begin{figure}[htb]
\vspace{9pt}
%\framebox[55mm]{\rule[-21mm]{0mm}{43mm}}
%\includegraphics[scale=0.6, angle=90]{G_vs_mu.eps}
%\includegraphics[scale=0.6]{G_vs_mu.eps}
\includegraphics[angle=270,width=17pc]{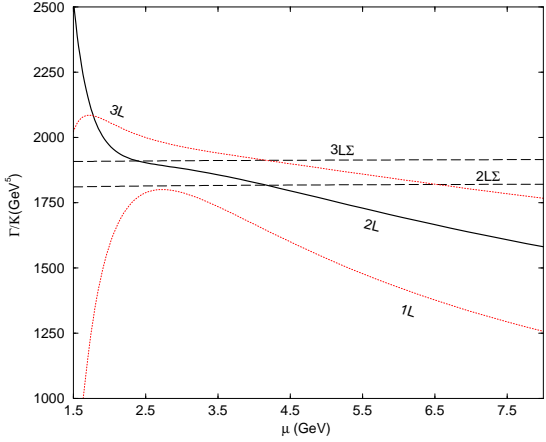}
\caption{Comparison of one-loop (1L), two-loop (2L), and three-loop
(3L) QCD $\overline{MS}$ corrections to the $b \rightarrow u \ell^- \overline{\nu}_\ell$
decay rate to corresponding rate expressions (2L$\Sigma$ and 3L$\Sigma$) obtained
via summation of all leading and subleading higher-order logarithm terms respectively accessible
from two-loop and three-loop QCD.  All curves are obtained using benchmark values
$\alpha_s (M_z) = 0.118, \; m_b (m_b) = 4.17$ GeV for the evolution of the QCD couplant and running
$b$-quark mass via the four-loop $\beta$-function and anomalous mass dimension.}
\label{Figure 1}
\end{figure}
Consequently, there is genuine ambiguity 
as to which value of $\mu$ is appropriate for theoretical predictions of $|V_{ub}|$  
from the inclusive semileptonic $B \rightarrow X_u$ decay rate, as well as concomitant 
theoretical uncertainty in such predictions.

The true decay rate $\Gamma$ is necessarily independent of $\mu$, an unphysical 
parameter introduced into perturbative QCD as a by-product of the regularization 
and the removal of infinities. Indeed the statement that the total derivative of   
$\Gamma$ with respect to $\mu$ must vanish, inclusive of implicit dependence of $\Gamma$ on $\mu$ through 
the QCD couplant $x(\mu)$ and running mass $m_b(\mu)$, leads directly to the 
renormalization-group equation (RGE)
\begin{equation}%3
\left[ 1 - 2\gamma_m (x) \right] \frac{\partial S}{\partial L} + \beta(x) \frac{\partial S}{\partial x} + 5\gamma_m S = 0,
\end{equation}
for the perturbative series
\begin{equation}%4
S\left[ x(\mu), L(\mu) \right] = \sum_{n=0}^\infty \sum_{k=0}^n T_{n,k} x^n L^k
\end{equation}
\begin{eqnarray}%5
T_{0,0} = 1, \; T_{1,0} = 4.25360, \; T_{1,1} = 5,\nonumber\\
T_{2,0} = 26.7848, T_{2,1} = 36.9902, T_{2,2} = 17.2917
\end{eqnarray}
within Eq. (1).  If we substitute the series (4) into the RGE (3), we obtain the 
following set of recursion relations by requiring that the aggregate coefficients 
of $x_n L^{n-1}, \; x_n L^{n-2}$, and $x_n L^{n-3}$ in the RGE (with
$\beta(x) = - \sum_{n=0}^\infty \beta_n x^{n+2}$ and anomalous mass dimension
$\gamma_m (x) = -\sum_{n=0}^\infty \gamma_n x^{n+1}$) respectively vanish:
\begin{equation}%6
n T_{n,n} - \left[ \beta_0 (n-1) + 5\gamma_0 \right] T_{n-1, n-1} = 0, \; 
 n \geq 1
\end{equation}
\begin{eqnarray}%7
0 = (n-1) T_{n,n-1} + 2\gamma_0 (n-1) T_{n-1, n-1}\nonumber\\
- \beta_0 (n-1) T_{n-1, n-2} - \beta_1 (n-2) T_{n-2, n-2}\nonumber\\
- 5\gamma_0 T_{n-1, n-2} - 5\gamma_1 T_{n-2, n-2} , \; n \geq 2
\end{eqnarray}
\begin{eqnarray}%8
0 & = & (n-2) T_{n, n-2} + 2\gamma_0 (n-2) T_{n-1, n-2}\nonumber\\
& + & 2\gamma_1 (n-2) T_{n-2, n-2} - \beta_0 (n-1) T_{n-1, n-3} \nonumber\\
& - & \beta_1 (n-2) T_{n-2, n-3} - \beta_2 (n-3) T_{n-3, n-3}  \nonumber\\
& - & 5\gamma_0 T_{n-1, n-3} -  5\gamma_1 T_{n-2, n-3} \nonumber\\
& - & 5\gamma_2 T_{n-3, n-3}, \; \; n \geq 3
\end{eqnarray}
The recursion relation (6) can be utilised to determine any 
coefficient $T_{n,n}$ in (4), given knowledge of $T_{0,0} (\equiv 1)$. 
Once all $T_{n,n}$ are known, the recursion relation (7) can be 
utilised to determine any coefficient $T_{n,n-1}$ from knowledge 
of $T_{1,0}$, as given by Eq. (5), and with this knowledge, the 
recursion relation (8) can be employed to determine any 
coefficient $T_{n,n-2}$ from knowledge of $T_{2,0}$ [Eq.(5)]. Upon 
rearrangement of the series (4) into the form
\begin{eqnarray}%9
S[x(\mu), L(\mu)] = \sum_{n=0}^\infty x^n S_n (x L), \nonumber\\
S_n (u) = \sum_{k=n}^\infty T_{k,k-n} u^{k-n}
\end{eqnarray}
we thus see that the leading three coefficients $S_0 (xL)$, $S_1 (xL)$, and $S_2 (xL)$ 
are fully determined by the recursion relations (6-8) and the values of
$T_{0,0}$, $T_{1,0}$, and $T_{2,0}$ already known from the two-loop calculation (1).  
The detailed mathematical evaluation of these coefficients $S_n (x L)$ is presented in ref. \cite{MRA02}, 
and leads to an optimal RG-improvement of the two-loop rate,
\begin{equation}%10
\Gamma_{2 L\Sigma} / \kappa = [m_b (\mu)]^5 [S_0 + S_1 x(\mu) + S_2 x^2 (\mu)]
\end{equation}
\begin{equation}%11
S_0 = \left[ 1 - \frac{23}{12} x(\mu) L(\mu) \right]^{-60/23}  
\end{equation}
\begin{eqnarray}%12
S_1 & = & -\frac{18655}{3174} \left[ 1 - \frac{23}{12} x(\mu) L(\mu) \right]
^{-60/23}\nonumber\\
& + & \left\{10.1310 + \frac{1020}{529}\log \left[1 - \frac{23}{12} x(\mu)
L(\mu) \right] \right\}\nonumber\\
& \times & \left[ 1 - \frac{23}{12} x(\mu) L(\mu) \right]^{-83/23}
\end{eqnarray}
\begin{eqnarray}%13
S_2 & = & 13.2231 \left[ 1-\frac{23}{12} x(\mu)L(\mu)\right]^{-60/23}\nonumber\\
& - & \left\{47.4897 + \frac{3171350}{279841} \right.\nonumber\\
& \times & \left. \log \left[1-\frac{23}{12}
x(\mu) L(\mu) \right] \right\}\nonumber\\
& \times & \left[1-\frac{23}{12} x(\mu)L(\mu)\right]^{-83/23} + \left\{61.0515 \right.\nonumber\\
&+ & 25.5973 \log \left[1-\frac{23}{12} x(\mu)L(\mu)\right]\nonumber\\
& + & \frac{719610}{279841} \left. \log^2 \left[1-\frac{23}{12} x(\mu)L(\mu)\right] \right\}\nonumber\\ 
& \times & \left[1-\frac{23}{12} x(\mu) L(\mu)\right]^{-106/23}
\end{eqnarray}
whose dependence on $\mu$ is virtually eliminated.  The $2L\Sigma$ curve in Fig. 1, 
as determined from Eq. (10), is almost flat over the range of $\mu$ indicated:  
$\Gamma_{2L\Sigma}/\kappa  = 1816 \pm 6$ GeV$^5$.

To obtain some control over 2L-order series truncation, we utilise an asymptotic 
Pad\'e-approximant determination of $T_{3,0} = 206$ \cite{MRA00} in conjunction 
with the RGE (3) to estimate the three-loop rate:
\begin{eqnarray}%14
\frac{\Gamma^{N^3 L}}{\mathcal{K}} & = & \left[m_b (\mu)\right]^5 \left\{ 1 +
(4.25360 + 5 L (\mu)) x(\mu) \right. \nonumber\\
& + & \left(26.7848 + 36.9902 L(\mu)\right.\nonumber\\
& + & \left. 17.2917 L^2 (\mu) \right) x^2(\mu) \nonumber\\
& + &  \left( 206 + 249.592 L(\mu) + 178.755 L^2 (\mu) \right. \nonumber\\
& + & \left. \left. 50.9144 L^3 (\mu) \right) x^3 (\mu) \right\}
\end{eqnarray}
The plot of this 3L rate in Fig. 1 indicates both a PMS maximum and an FAC point 
(at which the 3L term in Eq. (14) vanishes) at very low values of $\mu \; (\cong 1.8$ GeV $)$, 
after which the rate falls off with increasing $\mu$ somewhat less steeply 
than $\Gamma_{2L}/\kappa$. However, given knowledge of $T_{3,0}$, the RGE (3) 
may be used as above to calculate a recursion relation which determines all coefficients
$T_{n, n-3}$:
\begin{eqnarray}%15
0 & = & (n-3) \left[ T_{n, n-3} + 2\gamma_0 T_{n-1, \; n-3} \right. \nonumber\\
& + & \left.  2\gamma_1 T_{n-2, \; n-3} + 2\gamma_2 T_{n-3, \; n-3} \right] \nonumber\\
& - & \beta_0 (n-1) T_{n-1, \; n-4}- \beta_1 (n-2) T_{n-2, \; n-4}\nonumber\\
& - & \beta_2 (n-3) T_{n-3, \; n-4} -\beta_3 (n-4) T_{n-4, \; n-4} \nonumber\\
& - & 5\gamma_0 T_{n-1, \; n-4} -5\gamma_1 T_{n-2, \; n-4}\nonumber\\
& - & 5\gamma_2 T_{n-3,\; n-4} - 5\gamma_3 T_{n-4, \; n-4}, \; \; \; n \geq 4.\nonumber\\
\end{eqnarray}
This relation determines all coefficients within the Eq. (9) series expression for
$S_3 (xL)$ \cite{MRA02}:
\begin{eqnarray}%16
S_3 & = & -\frac{4.7895}{(1-\frac{23}{12} xL)^{60/23}}\nonumber\\
& + & \frac{[60.617 + 25.496 \log (1-\frac{23}{12} xL)]}{(1-\frac{23}{12} xL)^{83/23}} \nonumber\\
& + & \left[ -198.79-118.29 \log(1-\frac{23}{12} xL) \right. \nonumber\\
& - & \left. 15.114 \log^2 (1-\frac{23}{12} xL) \right] / (1-\frac{23}{12} xL)^{106/23} \nonumber\\
& + & \left[ 348.96+189.05 \log (1-\frac{23}{12} xL)\right.\nonumber\\
& + & 41.697 \log^2 (1-\frac{23}{12} xL)\nonumber\\
& + & \left.  2.9199 \log^3 (1 - \frac{23}{12} xL) \right] / (1-\frac{23}{12} xL)^{129/23}\nonumber\\
\end{eqnarray}

In Fig. 1, the 3L$\Sigma$ curve
$[m_b (\mu)]^5 [S_0 + S_1 x(\mu) + S_2 x^2 (\mu) + S_3 x^3 (\mu)]$
is plotted, based upon the estimated value $S_3(0) = T_{3,0} = 206$, and is seen 
to be even flatter than the 2L$\Sigma$ curve: $\Gamma_{3L\Sigma}/\kappa = 1912 \pm 4$ GeV$^5$
over the region of $\mu$ displayed.  This value is 5\% larger than the 2L$\Sigma$ rate, indicative of the 
truncation error from ignoring higher order terms.  Note that this estimate of 
truncation error is fully decoupled from renormalization scale dependence, 
which is virtually eliminated from both $2L\Sigma$ and (estimated) 3L$\Sigma$ rates. 
Curiously, $\Gamma_{2L\Sigma}$, the optimally RG-improved two-loop rate, is quite close 
to the PMS maximum of the one-loop rate $\Gamma_{1L}$, and that $\Gamma_{3L\Sigma}$, the optimally 
RG-improved three-loop rate, is quite close to the FAC prediction of the 
two-loop rate $\Gamma_{2L}$, suggesting that such benchmark points may anticipate 
higher order calculations.

We are grateful for hospitality from the KEK Theory Group, where this research was initiated, and for support from the Natural Sciences and Engineering Research Council of Canada.

\end{document}